\documentclass[10pt,a4paper,twoside]{article}
\usepackage{epsfig}
\usepackage{baltlat6}
\usepackage{array}
\usepackage{here}
\begin{document}
\ \
\vspace{0.5mm}
\setcounter{page}{209}

\titlehead{Baltic Astronomy, vol.\,23, 209--220, 2014}

\titleb{STOCHASTIC 2-D MODELS OF GALAXY DISK EVOLUTION.
THE GALAXY M33}

\begin{authorl}
\authorb{T. Mineikis}{1,2} and
\authorb{V. Vansevi\v{c}ius}{1,2}
\end{authorl}

\begin{addressl}
\addressb{1}{Vilnius University Observatory,
\v{C}iurlionio 29, Vilnius LT-03100, Lithuania;
vladas.vansevicius@ff.vu.lt}
\addressb{2}{Center for Physical Sciences and Technology,
Savanori\c{u} 231, Vilnius LT-02300, Lithuania;
tadas.mineikis@ftmc.lt}
\end{addressl}

\submitb{Received: 2014 November 20; accepted: 2014 December 23}

\begin{summary}
We have developed a fast numerical 2-D model of galaxy disk
evolution (resolved along the galaxy radius and azimuth) by adopting a scheme
of parameterized stochastic self-propagating star formation. We explore the
parameter space of the model and demonstrate its capability to reproduce 1-D
radial profiles of the galaxy M33: gas surface density, surface brightness in the
{\it i} and GALEX {\it FUV} passbands, and metallicity.
\end{summary}

\begin{keywords} galaxies:  evolution -- galaxies: individual (M\,33) \end{keywords}
\resthead{Stochastic 2-D models of galaxy disk evolution. M33}
{T. Mineikis, V. Vansevi\v{c}ius}

\sectionb{1}{INTRODUCTION}

Numerous recent resolved photometric and spectroscopic surveys of galaxies
allow us to explore a two-dimensional (2-D) structure of galaxy disks (resolved
along the galaxy radius and azimuth) in detail. Studies of disk structures have
proved them to be feature-rich (e.g., Bastian et al. 2007; Gieles et al. 2008;
Bastian et al. 2009; Sanchez et al. 2010). On the other hand, simulations of galaxy
disks with hydrodynamical models are computationally costly and dependent on
the parametrization of sub-grid physics and even on the implementation methods
(Scannapieco et al. 2012). Additionally, in the cases of late-type spirals fine tuning
of the models is needed (Ro\v{s}kar et al. 2014, and references therein). An increased
spatial resolution of the simulations could be the key to the problem, however,
this would make galaxy models computationally even more daunting (Guedes et
al. 2011).

In the past, there was an attempt to create fast semi-analytic 2-D models
of galactic disks, based on a stochastic self-propagating star formation (SPSF)
scenario (Seiden \& Gerola 1982, and references therein). The SPSF approach was
shown to be suitable to form flocculent spiral arms reminiscent of late-type spirals
(Gerola \& Seiden 1978). The properties of dwarf galaxies (Gerola et al. 1980),
as well as larger spirals (Feitzinger et al. 1981), were explained by employing
the SPSF scenario. Despite further developments of the grid-free SPSF variants
(Jungiewert \& Palou{\v s} 1994; Sleath \& Alexander 1995), galaxy modeling shifted
towards N-body and hydrodynamical approaches.

In this study we present an updated version of the 2-D stochastic galaxy disk model by Mineikis \& Vansevi\v{c}ius (2010). The most important new features of the
updated version are the implementation of the radial profile of gas accretion onto
the disk, based on the present-day mass distribution in the galaxy M33, and the
stellar ``dispersion'' across the disk. These features allow us to produce realistic
radial profiles of the surface brightness. We also improved the model by minimizing
a number of free parameters and designing it for fast generation of the results of
resolved stellar photometry. 

In order to verify the model, we explored an extensive grid of parameters by
comparing the model generated radial profiles of one-dimensional (1-D) galaxy
disk with the observed ones. For the study we used the well explored Local Group
galaxy M33. The relatively unperturbed stellar disk (Ferguson et al. 2007) and its
favorable inclination together with the small distance -- 840 kpc (Freedman et al.
2001), as well as the low Galactic extinction in its direction, makes this galaxy one
of the best laboratories for studies of the SPSF scenario. M33 has the characteristic
flocculent spiral arms inherent to the late-type disk galaxies. Although two spiral
arms are visible in the near infrared (Regan \& Vogel 1994), they do not dominate
the present-day star formation pattern, as can be seen on the far ultraviolet and
H$\alpha$ images (Thilker et al. 2005). The abundance of observational data makes the
M33 galaxy the best target to test our 2-D model.
 
The article is organized as follows. In Section 2 we describe the model, in
Section 3 the model parameters are calibrated for the galaxy M33, a discussion of
the simulation results is given in Section 4 and, finally, in Section 5 the conclusions
are presented.

\sectionb{2}{THE MODEL}

\subsectionb{2.1}{Disk geometry and time step}

The disk model is divided into $N_{\rm R}$ rings of equal width (Mineikis \& Vansevi\v{c}ius
2010). Each ring is subdivided into cells, which are the smallest elements of the
model. This subdivision follows the rule that the ring with the running number
$i$ has $6 \cdot i$ cells, producing a total of $3N_{\rm R} (N_{\rm R}-1) + 1$ cells in the disk. This
subdivision results in equal area cells, except for the central one, which is smaller
by a factor of 3/4. The physical size of the disk model is defined by the physical
size of cells and the number of rings.

We assume the size of cells to be comparable to the typical size of OB associ-
ations. According to the recent findings by Bastian et al. (2007), star formation
(SF) is a hierarchical scale-free process, highly dependent on the definitions. Nev-
ertheless, our model is not very sensitive to the adopted cell size, therefore, in this
study, a cell size is set to $d_{\rm C} = 100$ pc.

The model integration time step ($\Delta t_{\rm I}$) corresponds to the SF propagation time
across the cell. We assume that the SF propagation velocity is $v_{\rm SF} = 10$ km/s
(Feitzinger et al. 1981), i.e., it corresponds to the typical speed of sound in the
interstellar matter within a cell, therefore:
\vskip-2.2mm
\begin{equation}
\Delta t_{\rm I} = 10~{\rm Myr}\cdot \frac{d_{\rm C}}{100~{\rm pc}} \cdot
 \left( \frac{  v_{\rm SF}}{10~{\rm km/s} } \right)^{-1}.
\end{equation}

\pagebreak

\subsectionb{2.2}{Gas accretion}

Although mass accretion histories of dark matter (DM) halos are highly vari-
able (McBride et al. 2009), they tend, on average, to obey simple relations.
McBride et al. (2009) studied the Millennium simulation data and found a simple
empirical fit to the average DM accretion rate. The fit relates the mean accretion
rate of DM to a given halo mass at a given redshift. Using the higher resolution
Millennium II simulation, Fakhouri et al. (2010) confirmed the validity of the fit
and extended it to smaller halo masses. We use the results by Fakhouri et al.
(2010) to prescribe the DM and, correspondingly, baryonic mass (BM) build-up of
the model.

The gas falling into the DM halo is distributed in the thin disk. The radial
profile of the accretion is assumed to be a scaled version of the total BM radial
profile of the present-day galaxy disk:

\begin{equation}
A_{\rm G}(r,t)=\frac{B(r)}{2\pi \int_0^R B(r) r {\rm d} r}\cdot A_{\rm DM}(t) \cdot \beta ,
\end{equation}
where $A_{\rm G}(r,t)$ is the radial profile of gas accretion rate, $B(r)$ is the radial profile
of BM density in the galaxy disk, $A_{\rm DM}(t)$ is the rate of DM accretion on the
galaxy halo, and $\beta=\Omega_{\rm BM}/\Omega_{\rm DM}$ is the primordial ratio of BM to DM. Such a
definition of accretion guarantees the build-up of the present-day disk over the
galaxy's life-time. The metallicity of the accreted gas is set to $Z_{\rm A}=0.0001$.

\subsectionb{2.3}{Star formation}
The star formation process in the disk is modeled by discrete SF events occur-
ring stochastically in the cells. The cell can experience a SF event spontaneously
(probability  $P_{\rm S}$) or be triggered (probability $P_{\rm T}$).

$\bullet$ {\it Spontaneous SF probability $P_{\rm S}$}: this parameter defines SF events occurring
stochastically in the cells without any external influence. The spontaneous SF
sustains SF activity in the disk. In this study we set $P_{\rm S}$ to be very small, i.e., $\sim$1
SF event per model integration time-step $\Delta t_{\rm I}$ over the whole disk. Additionally,
we assume that spontaneous SF events are more probable in the cells possessing
higher gas density: $P_{\rm S} \propto \Sigma_{{\rm G}}^{2}$, where $\Sigma_{\rm G}$ is a surface gas density within a cell.

$\bullet$ {\it Triggered SF probability $P_{\rm T}$}:
this parameter represents a chain of complex
processes in the cell's interstellar medium (molecular cloud) after the SF event
has occurred. A molecular cloud will undergo imminent disruption by an energetic
feedback of stellar winds, expanding H\ II zones and supernovae explosions. Despite
a negative SF feedback locally (on a scale of cell), SF could be triggered on a larger
scale (i.e., in neighboring cells). For the cell $i$, which experiences a SF event in
a time step $t_k$, the neighboring cell $j$, being in direct contact with the cell $i$, can
experience a SF event during the next time step $t_{k+1}$ with a probability of $P_{{\rm T},j}$.

At the start of galaxy simulation the SF in the disk is inhibited by the critical
gas surface density $\Sigma_{\rm C}$. If the gas surface density in a cell is below critical, the
probability of SF event is reduced by a factor of $\Sigma_{\rm G}/\Sigma_{\rm C}$. As the simulation evolves,
the gas density in the disk increases and SF becomes more probable starting from
the inner disk parts. In the outer disk, $\Sigma_{\rm G}$ remains small, compared to $\Sigma_{\rm C}$, for a
longer time and defines the extent of the SF disk. We note that $\Sigma_{\rm C}$ defines the
local density, i.e., not averaged azimuthally, as this parameter is usually derived
from observations.

%
\begin{table}[th]
\vbox{\tabcolsep=4pt
\parbox[c]{124mm}{\baselineskip=0pt
{\smallbf\ \ Table 1.}{ PEGASE-HR parameters used to generate SSPs.\lstrut}}
\begin{tabular}{l l l}      
\hline              
Parameter &  Value  & Reference \\ 
\hline                       
Stellar library              & low-resolution & Le Borgne et al. (2004) \\
Initial mass function        & corrected for binnaries  & Kroupa (2002) \\
Fraction of close binaries	 & 0.05 & default PEGASE-HR value \\
Ejecta of massive stars	     & type B & Woosley \& Weaver (1995) \\
Nebular emission             & true & PEGASE-HR value  \\
\hline
\end{tabular}
}
\vskip-2mm
\end{table}

During a SF event, a fraction of gas available in a cell is converted to stars.
This fraction is called the SF efficiency, SFE:

\begin{equation}
{\rm SFE} = \epsilon \cdot \left( \frac{ \Sigma_{\rm G}}{10~M_{\odot}/{\rm pc^2}} \right)^{\alpha} ,
\end{equation}
where $\epsilon$ and $\alpha$ are free parameters. In order to avoid generating unrealistic stellar populations of extremely low mass or SFE exceeding 100\%, we set the lower
and upper SFE cut-offs to 0.05\% and 50\%, respectively. Additionally, if the stel-
lar population, born in a cell, is less massive than $100\,M_{\odot}$, we assume that the
starburst does not trigger neighboring cells because the population is, on average,
lacking supernovae.

\subsectionb{2.4}{The evolution of cells}

Each stellar population, formed in a cell during a SF event, could be well
represented by a simple stellar population (SSP). To track the entire SF history of
a cell, the value of mass of each particular stellar population is stored in a 2-D (age
vs. metallicity) array, $S$. The step in age, $t_{\rm S}$, can be set in accordance to the size
of a cell (10 Myr throughout the paper). The step in metallicity ($0.1$--$0.3$ dex) is
predefined by the used dataset of interpolated isochrones. To track the evolution
of SSPs self-consistently, we employ the package PEGASE-HR (Le Borgne et al.
2004; see Table 1 for the parameters used).

The stars from any SSP, formed in a cell, can move to the neighboring cell,
therefore, arrays $S_j$ of the neighboring cells have to be modified accordingly. Stellar
spread to neighboring cells is implemented as a ``dispersion'' process. The evolution
of the stellar content in the cell $S_i$ is given by

\begin{equation}
\frac{\Delta S_{i}}{\Delta t} = \Psi_{i} - D_{\rm S} \cdot \sum_{j} \lambda_{i,j} \cdot  \left( S_{i}-S_{j} \right),
\end{equation}
where $\Psi_i$ is the mass of a newly formed SSP, $D_{\rm S}$ is the ``dispersion'' constant, $\lambda_{i,j}$
is the length of borderline between cell $i$ and its neighbors $j$. To prevent star flow
anisotropies near the center of galaxy, which occur due to change in the length
ratios of the cell sides, we correct corresponding $\lambda$ to keep star flows through each
side of the cell equal.

The evolution of gas content in the cell $G_i$ is given taking into account SSP evolution:
\begin{equation}
\frac{\Delta G_{i}}{\Delta t}=-\Psi_i+A_i -\sum_{j} F_{i,j} + \sum_k \sum_l ( S_{i} \circ R)_{k,l},
\end{equation}
where $A_i$ is the accreted gas mass, $F_{i,j}$ are the gas mass flows (see Section 2.5)
between cell $i$ and neighboring cells $j$, and $R$ is the array representing the mass
fraction of stellar populations returned to gaseous content, i.e., the last term\footnote{Also known as Hadamard (or Schur) product, for two matrices of the same dimension, $(B\circ C)_{k,l}=B_{k,l}\cdot C_{k,l}$ \label{refnote}}
represents the total gas mass returned to gas pool of the cell $i$ by stellar populations
evolving within this cell (the indices $k$ and $l$ denote the age and metallicity).

The metallicity ($Z_i$) evolution of the gas content in the \textit{i}-th cell is given by

\begin{equation}
\frac{\Delta (Z_{i}\cdot G_{i})}{\Delta t}=-\Psi_i \cdot Z_{i}+A_i\cdot Z_{\rm A} -\sum_j F_{i,j} \cdot Z_j + \sum_k \sum_l ( S_i \circ Z )_{k,l},
\end{equation}
where $Z_{\rm A}$ represents the metallicity of the accreting gas, $j$ denotes the neighboring
cells, and $Z$ is the array representing yields of metals returned to the gas pool of
the i-th cell by stellar populations evolving within this cell.

\subsectionb{2.5}{Gas flows}

Due to the small size ($\sim100\times100$ pc) of cells covering the galaxy disk, gas
masses are expected to flow beyond the cells in one time step. The main structures
causing gas movement are superbubbles inflated by SF feedback. Following the
formulation by Castor et al. (1975), on time-scale of 10 Myr superbubbles reach $\sim100$ pc in size for a typical gas density of $20\,M_{\odot}/{\rm pc^2}$ and SFE values of a few
percent. Therefore, a cell experiencing a SF event on a time-scale of 10 Myr will
be filled with hot tenuous gas, causing a large part of the gas to move out of the
cell.

After 40 Myr, when supernovae vanish and the superbubble inflated cavity
cools down, the gas returns to the cell. The force driving the gas to refill the cell
is the random motion of H\,I gas clouds having typical velocities of $\sim10$ km/s (see,
e.g., Corbelli \& Schneider (1997) for the galaxy M33). Based on the arguments by
Hunter \& Gallagher (1990), we estimate the refill time-scale $\tau=d_{\rm C}/(2\cdot10~{\rm km/s})\simeq 50$ Myr.

We implemented both types of gas flows in the galaxy disk model: expulsion
by a SF event and refilling.

$\bullet$ {\it Gas expulsion}: this is implemented simply by moving gas from the cell $i$,
experiencing a SF event, to the neighboring cells $j$. The gas flows only to those
neighbors which have had no a SF event for the past 40 Myr:
 
\begin{equation}
F_{i,j} =  \lambda_{i,j} \cdot \left\{ 
  \begin{array}{l l}
     0      & \Delta_{{\rm SF},j}\le 40\,{\rm Myr} \\
     G_{i}  & \Delta_{{\rm SF},j} > 40\,{\rm Myr}, \\
  \end{array} \right.
\end{equation}
where gas flows between cells $F_{i,j}$ depend on the length, $\lambda_{i,j}$, of the borderline,
and $G_i$ is the gas mass remaining in the cell $i$ after the SF event. The flow occurs
within one time step after the SF event.

$\bullet$ {\it Gas refilling}: this is implemented under the assumption that equilibrium
gas distribution in the galaxy follows the present-day baryonic matter distribution
$B(r)$, i.e., any deviation from it, ($G(r_i)/G(r_j)\neq B(r_i)/B(r_j)$), generates gas flows
between cells on the time-scale $\tau$:
\begin{equation}
F_{i,j} =  \lambda_{i,j} \cdot \frac{1}{\tau}\cdot \frac{G(r_j)\cdot B(r_i)-G(r_i)\cdot B(r_j)}{B(r_i)+B(r_j)}.
\end{equation}

%
\begin{table}[t]
\vbox{\tabcolsep=12pt
\parbox[c]{124mm}{\baselineskip=0pt
{\smallbf\ \ Table 2.}{ The parameters of the 2-D galaxy model.\lstrut}}
\label{table:ModelParams}
\begin{tabular}{l l l}   
\hline                 
Parameter & Notation &  Value  \\
\hline                       
Fixed parameters \\
\hline
Age				  & --  		 & 13 Gyr \\
Baryonic mass	  & -- 	         & $1.1\cdot10^{10}\,M_{\odot}$ \\
Disk radius		  & --           & 12 kpc       \\
Disk rotation			  & --      	 & Corbelli \& Salucci (2007) \\
Cell size 			      & $d_{\rm C}$   & 100 pc    \\
SF propagation speed      & $v_{\rm SF}$ & 10 km/s    \\
Accretion rate			  & $A_{\rm DM}(t)$       & Fakhouri et al. (2010) \\
Critical gas surface
  density                 &$\Sigma_{\rm C}$          & $8\,M_{\odot}/{\rm pc^2}$  \\  
Gas refilling time-scale		  & $\tau$  & 50 Myr      \\
Star ``dispersion'' constant			  & $D_{\rm S}$ & 200 ${\rm pc^2/Myr}$      \\
Spontaneous SF probability& $P_{\rm S}$    &  1 SF region   per $\Delta t_{\rm I}$ \\    
\hline                       
Varied parameters \\
\hline
Triggered SF probability  & $P_{\rm T}$            & 0.28-0.44      \\
SF efficiency &$\epsilon$ & (0.018-5.7)$\cdot 10^{-2}$  \\
SF power index&$\alpha$    & 1.5-2.5 \\
\hline
\end{tabular}
}
\vskip6mm
\end{table}

\sectionb{3}{THE MODEL SETUP FOR M33}   

The model parameters used for the M33 galaxy simulation are given in Table~2.
The model disk (radius of 12 kpc) comfortably engulfs the star forming disk of M33.
The presence of the RR Lyr variables in M33 (Pritzl et al. 2011 and references
therein) implies its old age, therefore, the galaxy formation was assumed to start
13 Gyr ago. The critical gas density for SF ($\Sigma_{\rm C}$) was suggested to be in the range
3--10 $M_{\odot}/{\rm pc^2}$ (Schaye 2004). We have found that lower end values do not change
the simulation results significantly and adopted $\Sigma_{\rm C}=8\,M_{\odot}/{\rm pc^2}$. The 1-D radial
profiles are not very sensitive to the stellar ``dispersion'' parameter, therefore, it
was set to $200~{\rm pc^2/Myr}$. The effects of this ``dispersion'' parameter will be analyzed
in the subsequent paper devoted to the study of properties of 2-D galaxy models.

An assumption was made that the radial profile of gas accretion is a scaled
version of the radial profile of total baryonic mass (gaseous and stellar) in the
galaxy. To avoid computationally expensive iterative fitting of baryonic mass
for each model, we derived the radial profile by performing the decomposition of
the galaxy's rotation curve. The observational data of the rotation curve were
taken from Corbelli \& Salucci (2007). The rotation curve was decomposed into
four mass components: gaseous disk, stellar nucleus, exponential stellar disk and
pseudo-isothermal DM halo.

The contribution of the gas disk to the rotation curve was calculated by co-adding H$\,$I and H$_2$ data from Corbelli \& Salucci (2000) and Heyer et al. (2004),
respectively. Helium mass fraction was taken into account by multiplying hydrogen
gas mass by a factor of 1.33. Corbelli (2003) has established that the mass of the
M33 stellar nucleus is in the interval $(0.3-8)\cdot 10^8\,M_{\odot}$, therefore, in order to narrow
a free-parameter space, we assume that the stellar nucleus is a point source with
a mass of $10^8\,M_{\odot}$. To avoid the effects of the extended nucleus, we did not fit the
innermost 0.5 kpc part of the rotation curve.

\begin{figure}[!th]
\begin{center}
\makebox[\textwidth]{\includegraphics[width=100mm]{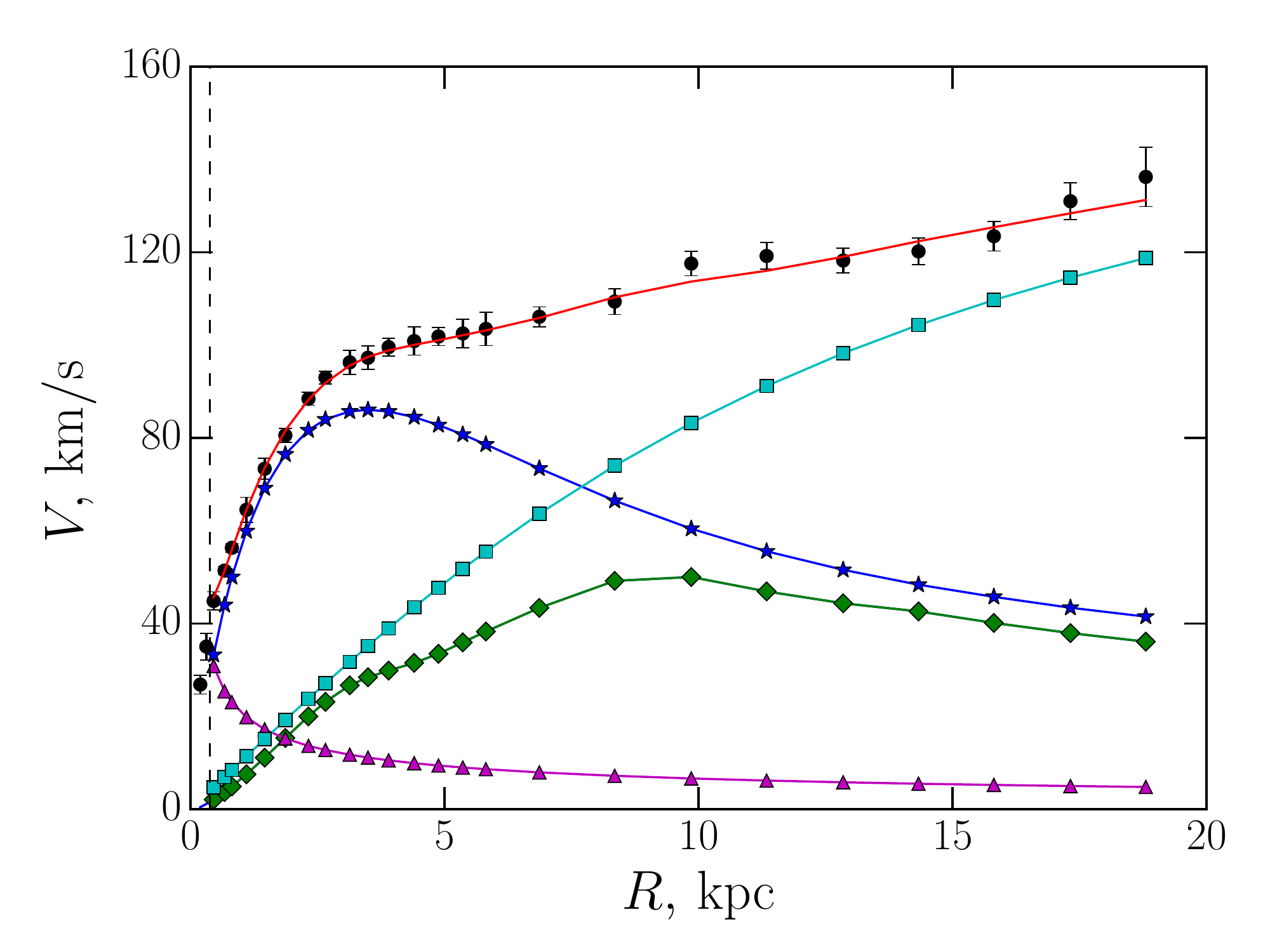}}
\end{center}
\vskip-7mm
\captionb{1}{Circular speed versus radius for the galaxy M33. The black dots with error
bars indicate the observed rotation curve (Corbelli \& Salucci 2007). The red line is the
best-fitting model rotation curve (see the text for the model description). Individual
contributions from the components are also shown: pseudo-isothermal DM halo (cyan
squares), exponential stellar disk (blue stars), gaseous disk (green diamonds) and stellar
nucleus (magenta triangles).}
\label{fig:rot}
\vskip-2mm
\end{figure}

We fitted the following two components:
\begin{itemize}

\item a pseudo-isothermal DM halo with the core radius $R_c$ and the asymptotic
rotation velocity $V_{R}(\infty)$, e.g., de Blok et al. (2008);
\item an exponential stellar disk with the central surface density $\Sigma_{\rm S}$ and the scale
length $R_e$. The stellar disk was assumed to have a scale height of 100 pc.
\end{itemize}

Dynamical contributions of gaseous and stellar disk components were computed
using the task \textit{rotmod} within the software package GIPSY (van der Hulst et al.
1992) which implements the method of Cesartano (1983). For each pair of the
parameters describing the stellar disk, $\Sigma_{\rm S}$ \& $R_e$, we minimized $\chi^2$ by fitting the
DM radial profile.

The best fitted rotation curve with ${\it R_e}=1.57$\,kpc, $\Sigma_{\rm S}=470\,M_{\odot}/{\rm pc^2}$, $R_c=10.2$\,kpc, and $V_{R}(\infty)=184$\,km/s is shown in Fig.~1. The derived scale length
of stellar disk is in agreement with the values derived photometrically, 1.4\,kpc in
the \textit{K}-passband for the inner 4\,kpc of M33 (Regan \& Vogel 1994) and $1.5$--$1.6$\,kpc
in the Spitzer passbands at 3.4 and 4.5 $\mu m$ (Verley et al. 2009). The best-fitted
stellar and gaseous disks were used to define the radial gas accretion profile of
M33.

The grid of models was computed by varying the main SF parameters: $\epsilon$, $\alpha$
and $P_{\rm T}$. The grid is composed of $17$ linear steps in $P_{\rm T}$, 5 linear steps in $\alpha$ and
11 logarithmic steps in $\epsilon$. In order to reduce stochastic effects, every model is
presented as an average of six snapshots spaced by 200\,Myr during the last 1\,Gyr.

\begin{figure*}[ht]
\includegraphics[width=\textwidth]{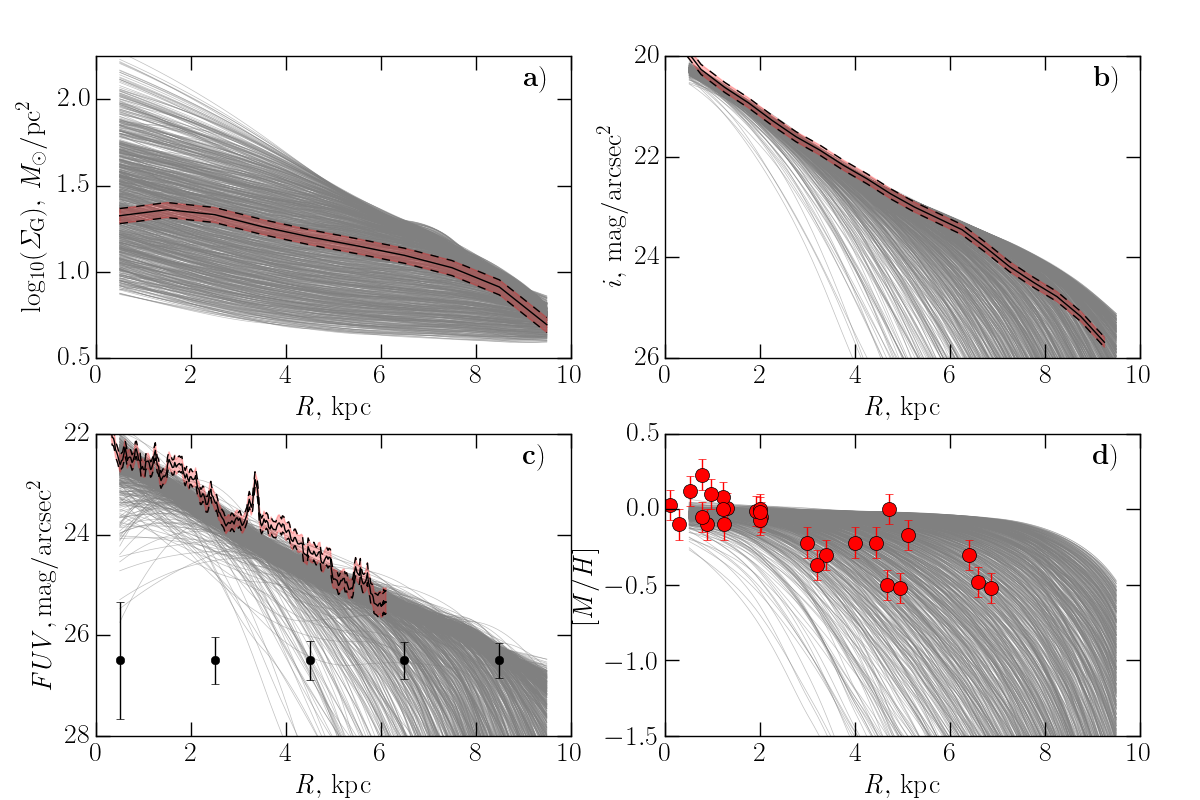}
\captionb{2}{Comparison of the observational data on the galaxy M33 with the models.
The gas surface density (panel a) is computed by co-adding the radial profiles of H\,I
(Corbelli \& Salucci 2000) and H$_2$ (Heyer et al. 2004) gas surface density. The surface
brightness radial profile in the $i$ passband (panel b) (Ferguson et. al. 2007) is corrected
for internal extinction using the radial extinction profile by Munoz-Mateos et al. (2007)
and assuming the LMC-like extinction law (Gordon et al. 2003). The surface brightness
radial profile in GALEX \textit{FUV} passband (panel c) (Munoz-Mateos et al. 2007) is also
corrected for internal extinction. The metallicity measurements of blue supergiant stars
(panel d) are taken from Urbaneja et al. (2005) and U et al. (2009). All radial profiles of
observed surface brightness are de-projected adopting 54$^\circ$ for the galaxy disk inclination.}
\label{fig:allmod}
\vskip-8mm
\end{figure*}

\sectionb{4}{RESULTS}

Comparisons of the models with observational data for the galaxy M33 are
presented in Fig.~2. The models are compared with 1-D radial profiles of: gas
surface density (H\,I from Corbelli \& Salucci 2000, H$_2$ from Heyer et al. 2004),
surface brightness in $i$ (Ferguson et al. 2007) and GALEX \textit{FUV} (Munoz-Mateos et al. 2007) passbands, and metallicity (Urbaneja et al. 2005; U et al. 2009).

The gas surface density radial profile derived from observations constrains
strongly the model parameter space due to large variation of the model gas density
radial profiles, especially in the inner parts of the galaxy (Fig.\,2, panel a).

The surface brightness radial profile in the $i$ passband constrains well the stellar
mass of the models. However, we set the same total mass for all models, thus there
are no significant variations in the $i$ passband radial profiles, except in the outer
parts of the galaxy, where model radial profiles span a wide range (Fig.\,2, panel~b).

The GALEX \textit{FUV} passband surface-brightness radial profile constrains well
the SF rate along the model galaxy radius (Fig.\,2, panel c). However, this profile
is not very sensitive to the model parameters since the accretion time-scale exceeds
the gas consumption time-scale, i.e., SF is gas-accretion regulated (Elmegreen 2014
and references therein).

\begin{figure}[H]
\includegraphics[width=130mm]{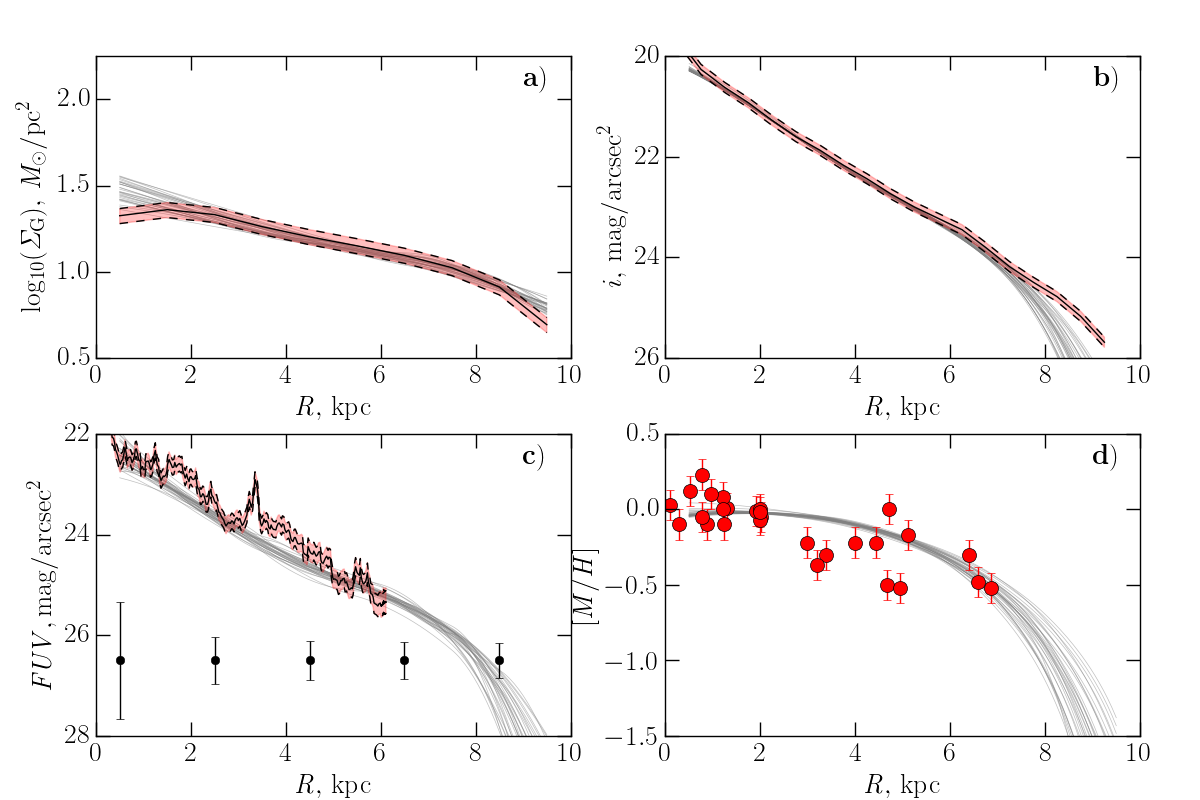}
\captionb{3}{Same as Fig.~2, but only models with r.m.s. deviations from the observed
radial profiles (gas surface density and $i$ passband) smaller than 10\% and in the radial
distance range 2--7\,kpc are shown.}
\includegraphics[width=130mm]{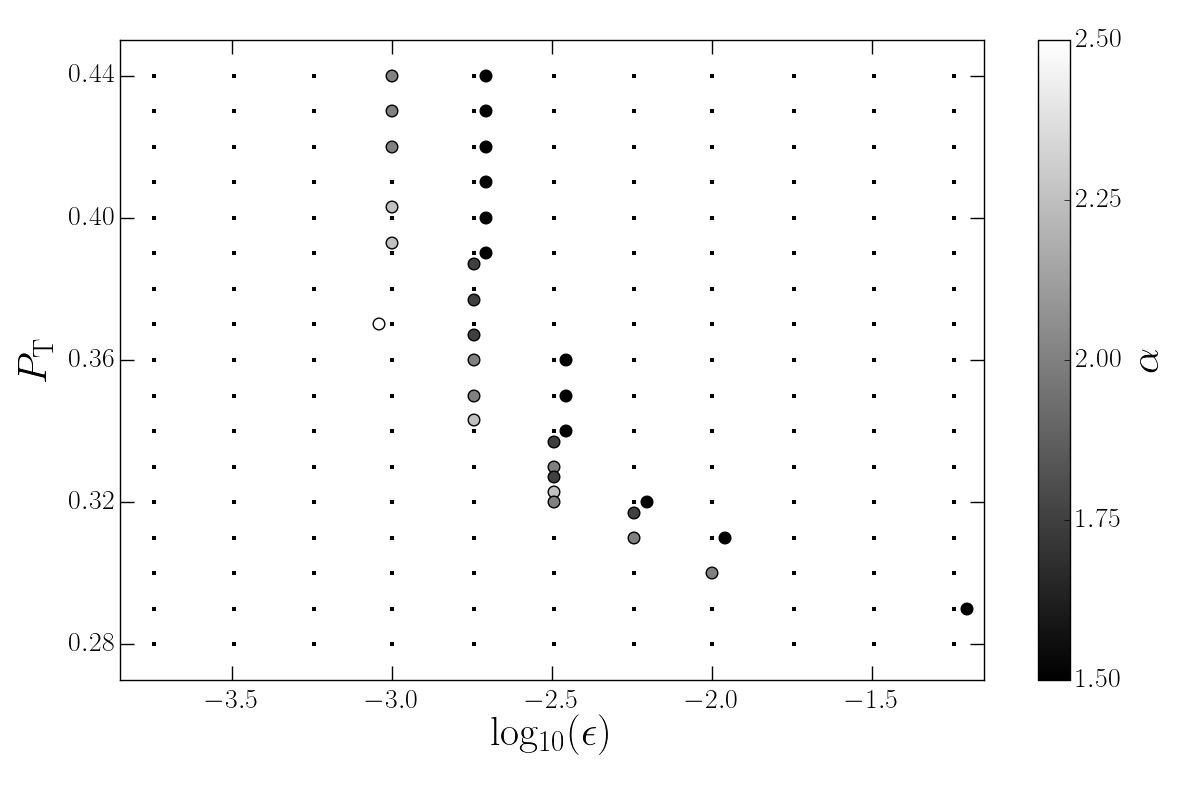}
\captionb{4}{3-D parameter space of the models shown in Fig.~3. The gray-shaded circles
indicate the values of $\alpha$ parameter. The small black points at each node of the model
grid indicate a fully explored extent of the parameter space.}
\end{figure}

\begin{figure}[H]
\includegraphics[width=130mm]{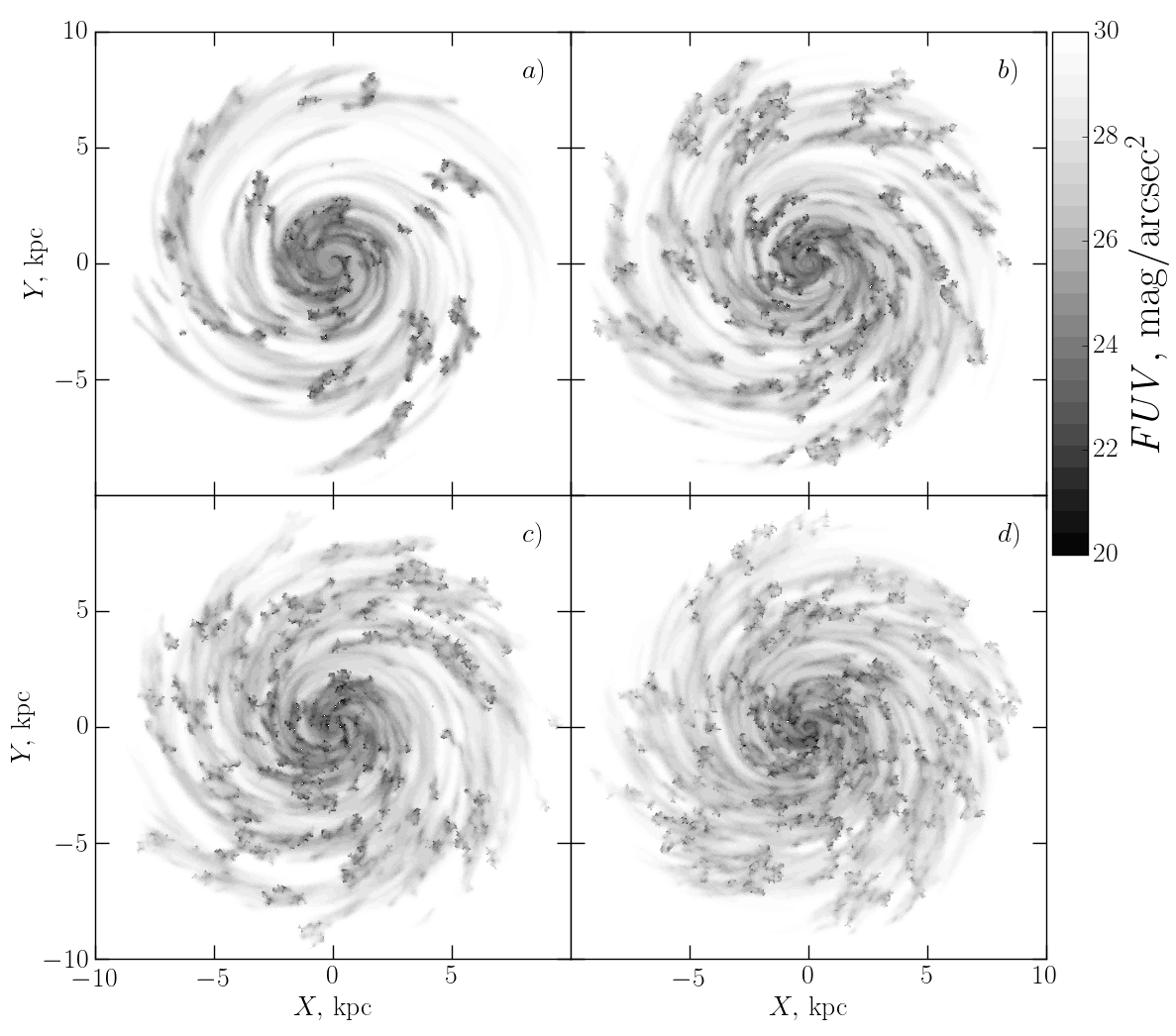}
\captionb{5}{The galaxy models from the ``degeneracy valley'' (Fig. 4) in the GALEX \textit{FUV} passband: (a) $P_{\rm T}=0.30,\,\epsilon=1\%$; (b) $P_{\rm T}=0.31,\,\epsilon=0.6\%$; (c) $P_{\rm T}=0.32,\,\epsilon=0.3\%$; (d) $P_{\rm T}=0.34,\,\epsilon=0.2\%$. For all models, $\alpha=2$. The models possess very similar 1-D radial
profiles, but they demonstrate different 2-D patterns of the young SF regions.}
\end{figure}

The metallicity radial profiles (Fig.\,2, panel d) are sensitive to the model parameters, however, they are of limited use because of large scatter in the observational
data.

Therefore, to further constrain the model parameter space we used only the
gas surface density and the $i$ passband radial profiles. For each model we calculated the r.m.s. deviations from both observed radial profiles in the most reliable
(for models) radial distance range, i.e., $2$--$7$\,kpc. The inner region of the model
galaxy has an increasingly anisotropic grid and the outer parts are affected by the
uncertainty in the critical gas surface density, $\Sigma_{\rm C}$. In Fig.\,3 we show the models
selected by r.m.s. deviations from the observed radial profiles (gas surface density
and $i$ passband) being less than 10\%.

The parameter space of the selected models is shown in Fig.\,4. A prominent
feature of this figure is the ``degeneracy valley'' -- the region in the parameter space
where reside the models selected by the 1-D radial profile procedure described
above. The ``degeneracy valley'' is seen because the 1-D radial profiles are derived
by convolving the 2-D distributions of SF regions, e.g., SF in the model galaxy
being smooth and inefficient or patchy and efficient would produce similar 1-D
radial profiles.

This effect is shown in Fig.~5, where four different models taken from the
``degeneracy valley'' of Fig.~4 are plotted. Fig.~5 illustrates that, by increasing $P_{\rm T}$
and decreasing $\epsilon$ values, SF becomes more smooth. It is challenging to break those
degeneracies by using only 1-D observed radial profiles, however, feature-rich 2-D model structures (Fig.~5) imply the possibility to solve this problem by employing
spatially resolved observations of real galaxy disks.

\sectionb{5}{CONCLUSIONS}

We presented a 2-D model of galaxy disk evolution, based on the stochastic
self-propagating star formation process. The parameterized gas and star dynamics,
together with self-consistent chemical and photometric evolution of stellar popu-
lations, are implemented in the model. The galaxy build-up prescription is taken
from the N-body DM simulations by Fakhouri et al. (2010).

We applied the model for the interpretation of observational data on the M33
galaxy and successfully reproduced the 1-D radial profiles of the gas surface density, GALEX \textit{FUV} and $i$ passband surface brightness, and stellar metallicity. We
explored possible degeneracies of the model parameters by computing the extensive grid of models. We have found that observational data, averaged as 1-D
radial profiles, cannot fully characterize galaxy evolution. In order to break the
degeneracies, the analysis of 2-D SF region patterns is additionally required.

\thanks{This research was partly funded by the grant No. MIP-074/2013 from the Research Council of Lithuania.}

\References

\refb Bastian N., Ercolano B., Gieles M. et al.\ 2007, MNRAS, 379, 1302 
\refb Bastian N., Gieles M., Ercolano B., Gutermuth R.\ 2009, MNRAS, 392, 868
\refb Castor J., McCray R., Weaver R.\ 1975, ApJ, 200, L107
\refb Casertano S.\ 1983, MNRAS, 203, 735
\refb Corbelli E., Schneider S.~E.\ 1997, ApJ, 479, 244
\refb Corbelli E., Salucci P.\ 2000, MNRAS, 311, 441
\refb Corbelli E.\ 2003, MNRAS, 342, 199 
\refb Corbelli E., Salucci P.\ 2007, MNRAS, 374, 1051
\refb de Blok W.~J.~G., Walter F., Brinks E. et al.\ 2008, AJ, 136, 2648 
\refb Elmegreen B.~G.\ 2014, arXiv:1410.1075 
\refb Fakhouri O., Ma C.-P., Boylan-Kolchin M.\ 2010, MNRAS, 406, 2267 
\refb Feitzinger J.~V., Glassgold A.~E., Gerola H., Seiden P.~E.\ 1981, A\&A, 98, 371 
\refb Ferguson A., Irwin M., Chapman S. et al.\ 2007, \textit{Island Universes -- Structure and Evolution of Disk Galaxies}, Springer, p. 239
\refb Freedman W.~L., Madore B.~F., Gibson B.~K. et al.\ 2001, ApJ, 553, 47
\refb Gerola H., Seiden P.~E.\ 1978, ApJ, 223, 129 
\refb Gerola H., Seiden P.~E., Schulman L.~S.\ 1980, ApJ, 242, 517
\refb Gieles M., Bastian N., Ercolano B.\ 2008, MNRAS, 391, L93
\refb Gordon K.~D., Clayton G.~C., Misselt K.~A., Landolt A.~U., Wolff, M.~J.\ 2003, ApJ, 594, 279 
\refb Guedes J., Callegari S., Madau P., Mayer L.\ 2011, ApJ, 742, 76 
\refb Heyer M.~H., Corbelli E., Schneider S.~E., Young J.~S.\ 2004, ApJ, 602, 723
\refb Hunter D.~A., Gallagher J.~S. III 1990, ApJ, 362, 480 
\refb Jungwiert B., Palous J.\ 1994, A\&A, 287, 55 
\refb Kroupa P.\ 2002, Science, 295, 82 
\refb Le Borgne D., Rocca-Volmerange B., Prugniel P. et al.\ 2004, A\&A, 425, 881
\refb McBride J., Fakhouri O.,  Ma C.-P.\ 2009, MNRAS, 398, 1858 
\refb Mineikis T., Vansevi\v{c}ius V.\ 2010, Baltic Astronomy, 19, 111 
\refb Mu{\~n}oz-Mateos J.~C., Gil de Paz A., Boissier S. et al.\ 2007, ApJ, 658, 1006
\refb Pritzl B.~J., Olszewski E.~W., Saha, A., Venn K.~A., Skillman E.~D.\ 2011, AJ, 142, 198 
\refb Regan M.~W., Vogel S.~N.\ 1994, ApJ, 434, 536
\refb Ro\v{s}kar R., Teyssier R., Agertz O., Wetzstein M., Moore B.\ 2014, MNRAS, 444, 2837 
\refb S{\'a}nchez N., A{\~n}ez N., Alfaro E.~J., Crone Odekon M.\ 2010, ApJ, 720, 541
\refb Scannapieco C., Wadepuhl M., Parry O.~H. et al.\ 2012, MNRAS, 423, 1726
\refb Schaye J.\ 2004, ApJ, 609, 667 
\refb Seiden P.~E., Gerola H.\ 1982, Fundamentals of Cosmic Physics, 7, 241 
\refb Sleath J.~P., Alexander P.\ 1995, MNRAS, 275, 507 
\refb Thilker D.~A., Hoopes C.~G., Bianchi L. et al.\ 2005, ApJL, 619, L67
\refb Urbaneja M.~A., Herrero A., Kudritzki R.-P. et al.\ 2005, ApJ, 635, 311
\refb U V., Urbaneja M.~A., Kudritzki R.-P. et al.\ 2009, ApJ, 704, 1120
\refb van der Hulst J.~M., Terlouw J.~P., Begeman K.~G., Zwitser W., Roelfsema P.~R.\ 1992, Astronomical Data Analysis Software and Systems I, 25, 131
\refb Verley S., Corbelli E., Giovanardi C., Hunt L.~K.\ 2009, A\&A, 493, 453
\refb Woosley S.~E., Weaver T.~A.\ 1995, ApJS, 101, 181 

\end{document}